# Roles of intrinsic anisotropy and π band pairbreaking effects on critical currents in tilted c-axis MgB$_2$ films probed by magneto-optical and transport measurements

A.A. Polyanskii, F. Kametani, D. Abraimov
  *Applied Superconductivity Center, National High Magnetic Field Laboratory, Florida State University, Tallahassee, Florida 32310, USA*

A. Gurevich
  *Department of Physics, Old Dominion University, Norfolk, VA 23529, USA*

A. Yamamoto
  *Department of Applied Chemistry, The University of Tokyo, Tokyo 113-8656, Japan
  and JST-PRESTO, Saitama 332-0012, Japan*

I. Pallecchi, M. Putti
  *CNR-SPIN and University of Genova, 16148 Genova, Italy*

C. Zhuang, T. Tan, X.X. Xi
  *Department of Physics, Temple University, Philadelphia, PA 19122, USA*

**Abstract**

Investigations of MgB$_2$ and Fe-based superconductors in recent years have revealed many unusual effects of multiband superconductivity but manifestations of anisotropic multiband effects in the critical current density J$_c$ have not been addressed experimentally, mostly because of the difficulties to measure J$_c$ along the c-axis. To investigate the effect of very different intrinsic anisotropies of σ and π electron bands in MgB$_2$ on current transport, we grew epitaxial films with tilted c-axis (θ ~ 19.5º), which enabled us to measure the components of J$_c$ both along the ab-plane and the c-axis using magneto-optical and transport techniques. These measurements were combined with scanning and transmission electron microscopy, which revealed terraced steps on the surface of the c-axis tilted films. The measured field and temperature dependencies of the anisotropic J$_c$(H) show that J$_{c,L}$ parallel to the terraced steps is higher than J$_{c,T}$ perpendicular to the terraced steps, and J$_c$ of thinner films (50 nm) obtained from transport experiments at 0.1 T reaches ~10% of the depairing current density J$_d$ in the ab plane, while magneto-optical imaging revealed much higher J$_c$ at lower fields. To analyze the experimental data we developed a model of anisotropic vortex pinning which accounts for the observed behavior of J$_c$ in the c-axis tilted films and suggests that the apparent anisotropy of J$_c$ is affected by current pairbreaking effects in the weaker π band. Our results indicate that the out-of-plane current transport mediated by the π band could set the ultimate limit of J$_c$ in MgB$_2$ polycrystals.

# 1. Introduction

The discoveries of two-band superconductivity in $MgB_2$ and more recently of a large family of multiband Fe-based superconductors have offered opportunities to investigate and improve superconducting properties by selective tuning of impurity scattering in different bands. This approach has been used successfully to increase the upper critical field $H_{c2}$ in $MgB_2$ much more effectively than is possible in single band superconductors [1]. However, the effects of bands with very different anisotropic parameters on flux dynamics and pinning in multiband superconductors have not been systematically investigated. For instance, most previous works on $MgB_2$ have focused on the field and temperature dependencies of the critical current density $J_c(T,H)$ in polycrystalline $MgB_2$ or single crystals in the ab plane. In the former case the observed $J_c$ can be masked by nonuniform current flow between misoriented grains in polycrystals in which the global $J_c$ is not entirely determined by the *ab*-plane properties [2]. Measurements of the depairing current density $J_d$ in the ab plane have been performed by Kunchur et al. using a high-current pulse technique [3], but the behavior of $J_c$ along the c axis is still not well understood.

The two bands in $MgB_2$ have markedly different characteristics: the π band is almost isotropic and has a smaller energy gap $\Delta_\pi(0) \approx 2\text{-}2.3$ meV, while the principal, highly anisotropic σ band has a larger gap $\Delta_\sigma(0) \approx 7\text{-}7.2$ meV [4]. Because the critical current density along the c-axis in the nearly two-dimensional σ band is strongly reduced, it is expected that the in-plane $J_c$ is mostly dominated by the σ contribution, whereas the out-of-plane $J_c$ is limited by the weaker and nearly isotropic π contribution, which can be suppressed more easily by field and temperature. Thus, the weaker π band may determine the ultimate upper limit of the global critical current density in polycrystals in which the π band mediates current links between neighboring crystallites with misoriented c-axes while the in-plane current transport is provided by the σ band. This scenario is consistent with the fact that the observed $J_c$ values in $MgB_2$ polycrystals are typically an order of magnitude smaller than $J_c$ measured in thin films [2] even though grain boundaries in $MgB_2$ are not weak links [5]. Understanding the mechanisms of current transport in $MgB_2$ polycrystals thus requires measurements of $J_c$ along the c-axis, which are very difficult because of small sizes of available $MgB_2$ single crystals and thin films along the c-axis. This experimental problem can be addressed by investigating epitaxial films with a tilted c-axis in which both the in-plane and the out-of-plane components of $J_c$, can be measured, and role of each band in determining the temperature and field dependence of the global $J_c$ can in principle be identified.

Recent materials advances have resulted in extremely high critical current densities in epitaxial $MgB_2$ thin films, ranging from $\sim 10^7$ A/cm$^2$ up to $10^8$ A/cm$^2$ [6,7,8,9,10,11]. The highest observed values of $J_c$ are of the order of the depairing current density, $J_d \sim 10^8$ A/cm$^2$ [3] which contains contribution of both bands. Each band can carry the maximum current density $J_d^{(i)} \sim \phi_0/4\pi\mu_0\lambda^2\xi_i$ inversely proportional to the respective coherence length $\xi_i$, [3] where i = (σ, π), λ is the London penetration depth and $\phi_0$ is the magnetic flux quantum. For typical values of $\xi_\sigma \approx 10$ nm and $\xi_\pi \approx 50$ nm in clean $MgB_2$ [12,13], the partial contribution $J_d^{(\pi)}$ of the π band is about 5 times smaller than $J_d^{(\sigma)}$ of the σ band. Such big difference between $J_d^{(\pi)}$ and $J_d^{(\sigma)}$, along with the isotropic character of the π band versus the anisotropic σ band, is yet another motivation to explore the critical current density in c-axis tilted films.

In this work, we investigate the anisotropic behavior of the critical current density $J_c$ in MgB$_2$ films grown on c-axis tilted (211) MgO substrates, for which the c-axis is rotated by θ ~ 19.5$^0$ away from the film plane normal. Using magneto-optical (MO) imaging at low magnetic fields and transport measurements at high fields, we measured both $J_c$ in the ab plane for the c-axis films, and the anisotropic $J_c$ for the mixed ab-plane and c-axis transport in the tilted films in the temperature range from 6 to 30 K. The measurements of $J_c$ were combined with scanning and transmission electron microscopy of the microstructure of our films. We also propose a model of strong single-vortex pinning and current pairbreaking in the π band, which captures the observed behavior of $J_c$ in the c-axis tilted films. The method developed in this work can also be used to probe anisotropic pinning of vortices in unconventional multiband superconductors, for example Fe-based pnictides or chalcogenides. The paper is organized as follows. In section 2 we describe parameters of our films and the experimental methods. In section 3 we describe structural, morphological, transport and MO experimental results. In section 4 we introduce a theoretical model of $J_c$ in tilted c-axis films. The model is then used for the interpretation of our experimental data in section 5. In section 6 we draw our conclusions and discuss implications of our results.

**2. Film samples and experimental methods**

Our MgB$_2$ thin films were produced by hybrid physical-chemical vapor deposition (HPCVD) on (211) MgO and (111) MgO substrates, as described in [14,15]. The use of (211) MgO substrates results in the c-axis tilted growth of MgB$_2$ films, which exhibit high structural and morphological homogeneity and no intergrain effects in transport properties, as was shown in ref. [16]. The behavior of $J_c$ in c-axis tilted films was compared with c-axis oriented MgB$_2$ films grown on (111) MgO substrates. Films of thickness t = 50 nm, 100 nm and 200 nm were deposited on each substrate. Four-probe resistivity and transport $J_c$ measurements were performed on laser-patterned micro-bridges made of the 50 nm and 100 nm films for which the effect of c-axis tilt on transport behavior is most pronounced. Our MO experiments were done using a 5 μm thick Bi-doped garnet indicator film, placed directly onto the sample surface as described elsewhere [17,18,19,20]. The MO technique was used for the reconstruction of induced magnetization currents in the films. All films were deposited onto 5×5 mm$^2$ square substrates for easy analysis of the $J_c$ anisotropy from MO patterns. Scanning electron microscopy (SEM) was used to investigate the surface microstructure of the films [14], while the film nanostructures were examined in JEOL JEM2011 and ARM200cF transmission electron microscopes (TEM).

**3. Experimental results**

*3.1 Structural and microstructural characterizations*

Our X-ray analysis has shown that the films deposited onto (211) MgO and (111) MgO substrates were epitaxial [21,14]. The (111) flat films are c-axis oriented, while the (211) films have their c-axis tilted by θ ~ 19.5° with respect to the normal to the substrate. Fig. 1 shows SEM images of the surfaces of three c-axis tilted films with the thicknesses of 50 nm (a), 100 nm (b), and 200 nm (c). Despite variations in the surface morphology, which depends sensitively on the variations in the deposition conditions, the SEM images for all the film thicknesses reveal characteristic terraced structures formed by hexagonal platelets of $MgB_2$ grains. The lateral dimensions of the hexagonal $MgB_2$ platelets are of the order ~100 nm, and the lateral widths of slanted terrace steps vary from tens of nm, that is of the order of the size of the vortex core, to ~100 nm. The hexagonal-shaped grains and rugged terrace edges should provide strong surface pins to fix vortex ends, irrespective of the direction of the in-plane current. The 200 nm thick film has larger lateral grain sizes L and step heights as compared to thinner films, indicating coalescence of $MgB_2$ crystallites during film growth. Figs. 2a and 2b show cross-sectional TEM images of the 50 nm and 100 nm thick c-axis tilted films. Both in 50 nm and 100 nm films, planar defects (marked by arrows) parallel to the *ab*-plane of $MgB_2$ and tilted by ~19.5° from the film surface appear at the $MgB_2$/MgO interface. Both images show strain contrasts (moiré patterns) caused by the planar defects due to the lattice mismatch at the $MgB_2$/MgO interface. These planar defects and their strain contrast extend through the whole thickness of the 50 nm film, but disappear in the upper half of the 100 nm film, presumably due to relaxation of lattice distortions. The high resolution TEM (HR-TEM) images of Figs. 2c and 2d show the terraced step structure at the $MgB_2$/MgO interface in the 50 nm and 100 nm films, confirming the epitaxial growth of $MgB_2$ even on the tilted substrate. The diffraction patterns derived by the fast Fourier transform (FFT) of different regions in Figs. 2e-f show that these straight planar defects are subgrain boundaries across which the $MgB_2$ lattice rotates around its c-axis, with a rotation angle up to 2-3 degrees. Most of these subgrain boundaries are parallel to the *ab*-plane of the $MgB_2$ lattice. The average spacing between the subgrain boundaries is ~30 nm, so that they could be additional pinning centers particularly in 50 nm film. Figs. 2a and 2b also show that these planar defects are neither stacking faults nor anti-phase boundaries nor dense arrays of dislocations induced by tilt steps which have been often observed in c-axis tilted $YBa_2Cu_3O_{7-\delta}$ films [22].

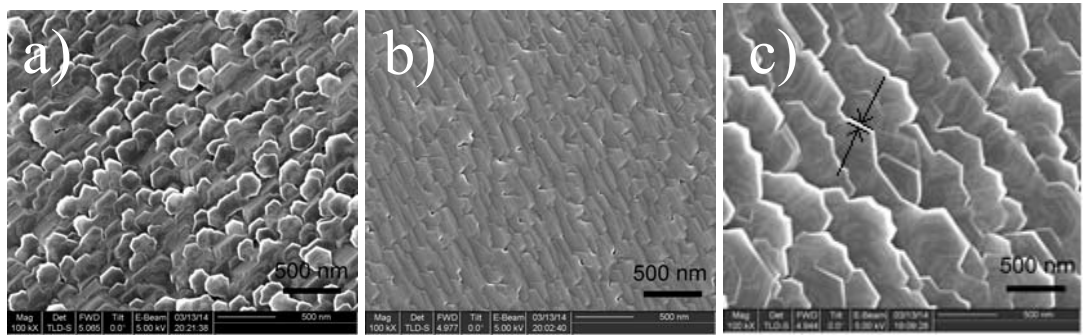

**Figure 1:** SEM images of terraced steps on the 19.5°c-axis tilted 50 nm (a), 100 nm (b) and 200nm (c) $MgB_2$ films. In panel (c), black arrows indicate the lateral width of slanted terrace steps, which is typically around tens of nm (see also Fig. 2a).

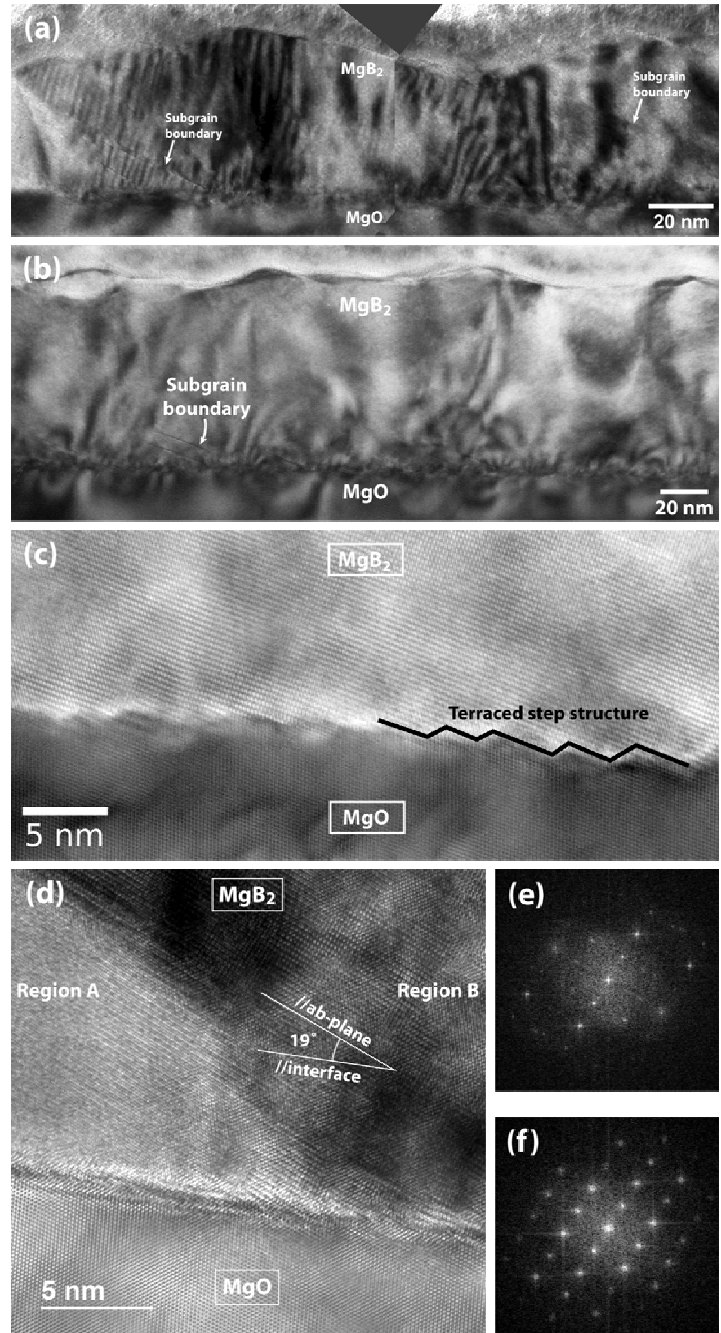

**Figure 2:** (a) Cross-sectional view of the 50 nm thick, 19.5° c-axis tilted $MgB_2$ film grown on the (211) MgO substrate. Planar subgrain boundaries appear at the interface and penetrate the whole film. (b) Cross-sectional view of the 100 nm tilted $MgB_2$ film on (211) MgO. Similar planar defects extend only to the middle of the film. (c) and (d) HR-TEM images near the $MgB_2$/MgO interface of the 100 nm and 50 nm tilted $MgB_2$ films, respectively. The ab-plane of $MgB_2$ tilts by ~19° with respect to the interface. (e) and (f) FFT derived diffraction patterns from regions A and B in (d), showing lattice rotation across the subgrain boundary.

*3.2 Resistivity measurements*

Resistivity measurements for each film were performed on 14 μm wide bridges, one of which is shown in the inset of Fig. 3. Bridges of the c–axis tilted films were laser-patterned parallel and perpendicular to the terraced steps in order to measure the normal state resistivities $\rho_L$ and $\rho_T$ for transport current parallel (L) and transverse (T) to the steps, respectively. Comparison of the observed $\rho_L$ and $\rho_T$ to the isotropic, in-plane resistivity $\rho$ of the c-axis oriented films is shown in Fig. 3, from which it follows that $\rho(T)$ almost overlaps with $\rho_L(T)$ for the c-axis tilted films for which current flows parallel to the steps. However the resistivity $\rho_T(T)$ of the c-axis tilted films measured with current flowing perpendicular to the steps, is noticeably larger than $\rho(T)$

To get insight into the results shown in Fig. 3, we first notice that the strongly temperature-dependent parts of $\rho_L(T)$, $\rho_T(T)$ and $\rho(T)$ caused by electron-phonon scattering are nearly identical both in 50 nm and 100 nm films, while the residual resistivities $\rho_0 = \rho(40 K)$ associated with impurity and surface scattering are different. The phonon-assisted drop in resistivity $\Delta\rho = \rho(300K) - \rho(40K)$ in our $MgB_2$ films is ~10 $\mu\Omega$ cm, which is slightly larger than but comparable with the intrinsic value $\Delta\rho \approx 4.3$ $\mu\Omega$ cm [23,24,25] observed on single crystals or strongly-coupled polycrystals [26] for which $\Delta\rho$ is dominated by current transport in the $\pi$ band at high temperature [27]. At lower temperatures, both bands contribute to the resistivity so $\rho$ and $\rho_L$ are nearly identical, because they both probe the *ab*-plane resistivity, while $\rho_T$ probes the resistivity both along the c axis and the ab planes. As long as the $\sigma$ band contribution along the c axis is small, the residual value of $\rho_T$ nearly coincides with the contribution of the $\pi$ band $\rho_{0T}$ and it is therefore larger than the residual resistivity of $\rho_0$ and $\rho_{0L}$, which probe both contributions or $\sigma$ and $\pi$ bands connected in parallel. We thus conclude that the resistivity in our c-axis tilted films is determined by intrinsic band parameters rather than by current-blocking grain boundaries or other extrinsic materials mechanisms which could cause the differences between $\rho_T$ and $\rho_L$. For instance, the thickness modulation due to the steps is much smaller than the thickness t, so the thickness can be regarded uniform both in the longitudinal and in the transverse configurations. In what follows we show how our c-axis tilted films can be used to study the intrinsic anisotropy of the critical current density in the superconducting state.

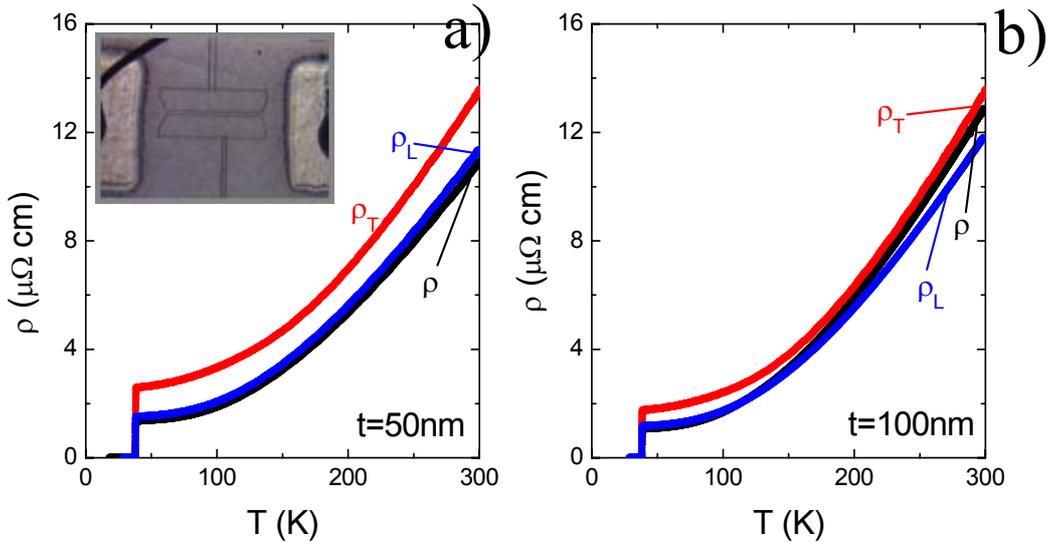

**Figure 3:** Temperature dependencies of resistivity for the c-oriented ($\rho$) and c-axis tilted ($\rho_L$ and $\rho_T$) $MgB_2$ films, patterned with micro-bridges oriented parallel ($\rho_L$) or perpendicular ($\rho_T$) to the terraced steps for the 50 nm film (a) and the 100 nm film (b). Inset shows one of the patterned 14 micron bridges used in the transport measurements.

*3.3 MO Imaging*

Fig. 4 shows the MO images of trapped flux patterns in the c-axis tilted and c-axis oriented films at 6 K after field cooling (FC) the films at 120 mT and then turning the field off. The bright diagonal lines (the so-called "current discontinuity d-lines") occur where the magnetization currents make 90° turns. Clearly, the Bean roof-top patterns in the c-axis oriented films (bottom panels) have a four-fold symmetry, while in the 50 nm and 100 nm c-axis tilted films (top panels) the roof slopes are steeper along the horizontal direction perpendicular to the terraced steps. These patterns indicate no $J_c$ anisotropy in the c-axis oriented films and anisotropy of $J_c$ in the 50 nm and 100 nm c-axis tilted films for which the critical current density flowing in the ab plane $J_{c,L}$ is larger than $J_{c,T}$ crossing the tilted ab planes. However, in the 200 nm c-axis tilted film the anisotropy of $J_c$ is significantly reduced. MO images of the 50 nm thick films shown in Figs. 4a and 4d reveal dendritic magnetic flux structures, which penetrate deeper along the terraced steps than across the steps. Such magnetic flux branching is characteristic of $MgB_2$ thin films with high $J_c$ at low temperature [28,29].

The dendritic flux instabilities disappear at temperatures above ≈ 9 K. For example, Fig. 5a shows a MO image taken at 10 K and 16 mT for ZFC c-axis tilted 50 nm film which is partly in the Meissner state. Here the bright regions of partial flux penetration correspond to higher values of the normal field component $B_z(x,y)$, while the Meissner state corresponds to the dark vortex-free central region. This MO image shows that the flux-penetrated regions are more extended along the surface steps than in the transverse direction, indicating that the critical current density along the steps $J_{c,L}$ is larger than $J_{c,T}$ perpendicular to the steps. Extraction of $J_c$ from the quantitative analysis of the widths of Meissner zone a(H) in anisotropic films is a complicated procedure due to the pillow-like shape of the penetrating flux front [30]. We did here a qualitative analysis of the $J_c$ anisotropy similar to that of ref. [18] using a simplified result of the Bean model, $a = w/\cosh(H/H_p)$ [30], where w is the sample width, H is the applied field, and $H_p = tJ_c/\pi$ The results are plotted in Fig. 5b for the films of different thicknesses t = 50, 100 and 200 nm. The so-obtained $J_{c,L}$ for the 50 nm film is extremely high, reaching almost $9 \cdot 10^7$ A/cm$^2$ at low temperature, while $J_{c,L}$ values for the 100 nm and 200 nm films are about $(4-5) \cdot 10^7$ A/cm$^2$. Fig. 5b also shows that the critical current densities decrease as the temperature increases for all films. Here $J_c$ of the c-oriented films is smaller than $J_{c,L}$ and $J_{c,T}$, while $J_{c,L}$ parallel to the steps is higher than $J_{c,T}$ across the steps. However, the 200 nm c-axis tilted film exhibit much weaker anisotropy at all temperatures, so that $J_{c,L} \approx J_{c,T} \approx J_c$

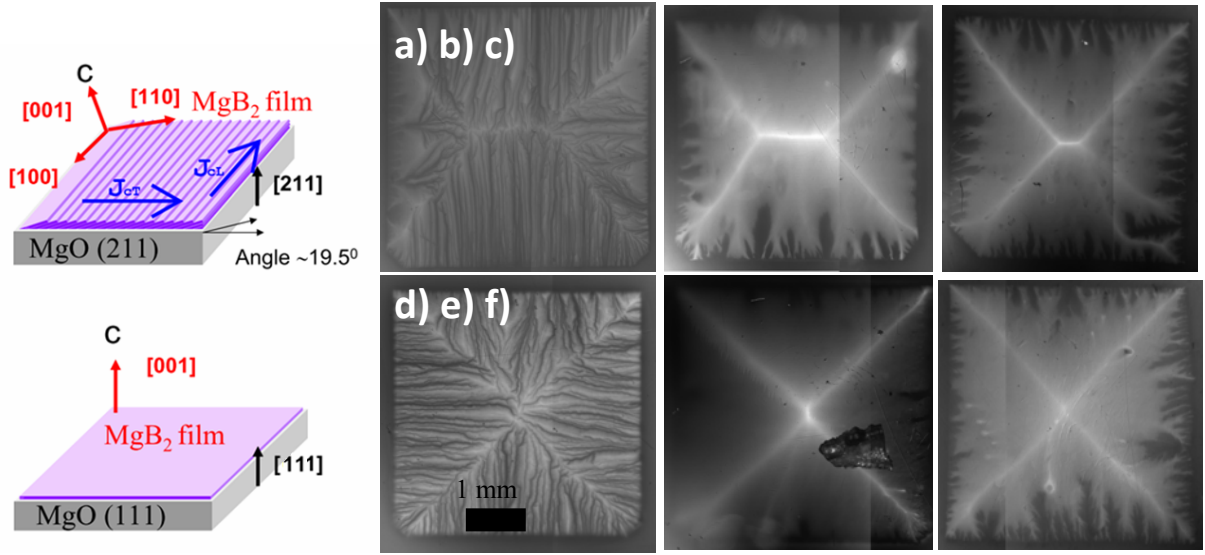

**Figure 4:** Magneto-optical images of trapped magnetic flux in the critical state of c-axis tilted (θ~19.5°) and flat samples field-cooled (FC) from the room temperature to 6 K in a magnetic field 120 mT which was then turned off. Top panels a, b and c show the c-axis tilted films of thickness 50, 100 and 200 nm, respectively, while bottom panels d, e and f show the c-axis oriented films of thickness 50, 100 and 200 nm. The terraced steps in the tilted films are parallel to the vertical edges of samples. Shown on the left are sketches of the c-axis tilted and c-axis oriented films and the definitions of $J_{c,T}$ and $J_{c,L}$.

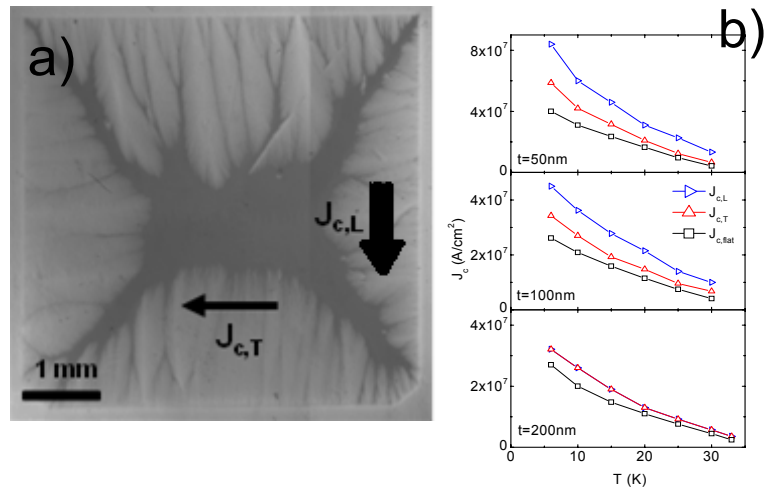

**Figure 5:** a) MO image of the c-axis tilted 50 nm film taken at 10 K in a field of 16 mT after cooling in zero field (ZFC). The critical currents parallel ($J_{c,L}$) and perpendicular ($J_{c,T}$) to the terraced steps (vertical direction) are indicated. The arrows show the direction of magnetization current in the ZFC regime and their different size indicates that $J_{c,L}>J_{c,T}$. b) Critical current versus temperature in the c-axis oriented and c-axis tilted films of thickness 50 nm, 100 nm and 200 nm, extracted from the width of the Meissner state in the vertical and horizontal directions. Note that for the 200 nm film $J_{c,L}$ and $J_{c,T}$ overlap.

Next, we extract the $J_{c,L}/J_{c,T}$ ratio from the MO images taken at fields higher than the field of full flux penetration, for which no corrections due to the complex shape of the Meissner region are necessary. The temperature dependence of $J_c$ anisotropy can be seen clearly in Figs. 6a and 6b, which show two MO images of the 50 nm c-axis tilted film in a fully penetrated magnetic flux critical state at T=10 K (a) and T=31 K (b). Here the length of the horizontal part of the roof pattern increases as temperature increases from 10 K to 31 K, departing from the four-fold symmetric pattern observed for the isotropic case. The square shape of our $MgB_2$ films allows us to evaluate quantitatively the current anisotropy using the anisotropic Bean model [30], which gives $J_{c,L}/J_{c,T}$ = 1.42 at T=10 K and $J_{c,L}/J_{c,T}$ = 2.2 at T=31 K. The temperature dependence of the anisotropy parameter $J_{c,L}/J_{c,T}$ in the c-axis tilted films of different thickness is plotted in Fig. 6c. For the 50 nm film, the ratio $J_{c,L}/J_{c,T}$ clearly increases as the temperature increases. A weaker but still visible increase of $J_{c,L}/J_{c,T}$ with temperature is also observed on the 100 nm film, while for the 200 nm c-axis tilted film the currents are almost isotropic at all temperatures. The same results for the temperature dependence of $J_{c,L}/J_{c,T}$ are also obtained from the plots shown in Fig. 5b.

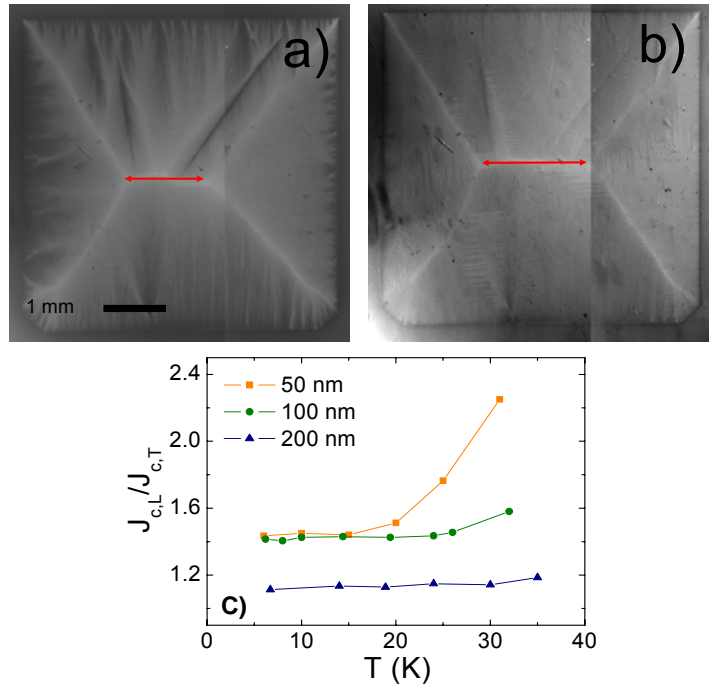

**Figure 6:** MO images of the c-axis tilted 50 nm film cooled down in a magnetic field of 120 mT which was then turned off at 10 K (a) and 31 K (b). The larger anisotropy of $J_c$ at T=31 K is apparent. Panel c shows temperature dependencies of the anisotropy parameter $J_{c,L}/J_{c,T}$ extracted from MO images of the c-axis tilted films of thickness 50 nm (squares), 100 nm (circles) and 200 nm (triangles).

The field dependence of the anisotropy parameter $J_{c,L}/J_{c,T}$ for the 50 nm c-axis tilted film was extracted from MO images measured in different fields. Fig. 7 shows examples of such images of ZFC tilted film at 52 mT (a) and 120 mT (b) at 6 K. As the field increases, the horizontal length of the Bean roof-top pattern increases yielding $J_{c,L}/J_{c,T}$=1.46 at 52 mT and 1.77 at 120 mT. Shown in Fig. 7c is the field dependence of $J_{c,L}/J_{c,T}$ measured at different temperatures. Here $J_{c,L}/J_{c,T}$ increases by 30-40% as the field increases from 52 mT to 120 mT.

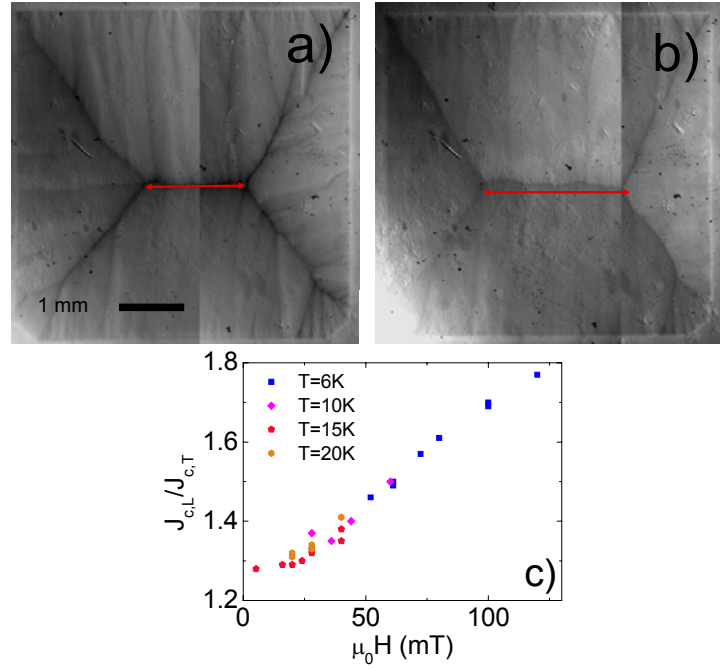

**Figure 7:** MO images of ZFC c-axis tilted film with t = 50 nm under the magnetic field of 52 mT (a) and 120 mT (b) at 6 K. Field dependencies of the anisotropy parameter $J_{c,L}/J_{c,T}$ extracted from MO images of this film under ZFC conditions at different fields and temperatures (c).

*3.4 Transport $J_c$ measurements*

In addition to the MO investigations described above, we also performed transport measurements of $J_{c,L}$ and $J_{c,T}$ using 14 μm bridges patterned parallel and perpendicular to the steps. The c-axis tilted films of thickness 50 and 100 nm were measured at 10, 20 and 30 K and different magnetic fields. Here $J_c$ was defined by the conventional electric field criterion, $E_c$=1 μV/cm in the region of magnetic fields below the irreversibility field H* where the exponent n(T,H) in the measured E-J characteristics $E=E_c(J/J_c)^n$ exceeds n > 3. Fig. 8 summarizes the temperature dependencies of transport $J_c$ measured in a perpendicular field of 0.1 T. The observed $J_c$ values are extremely high, exceeding $10^7$ A/cm$^2$, consistent with the MO data measured in the Meissner state at lower fields (see Fig. 5b). The transport $J_c$ values of the 50 nm films are systematically larger than those of the 100 nm films at 10 K and 20 K, in agreement with MO data, while at 30 K the trend is inverted, possibly because H becomes too close to the irreversibility field. These transport data also confirm the MO result that $J_{c,L}$ is larger than $J_{c,T}$ and both of them are larger than $J_c$

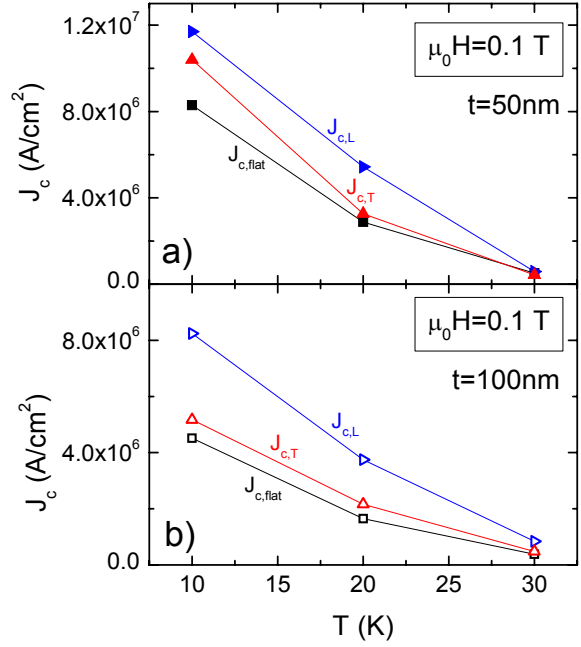

**Figure 8:** Temperature dependence of transport $J_c$ measured on the 50 nm (a) and 100 nm (b) c-axis tilted and c-axis oriented films in a perpendicular field of 0.1 T.

We also measured transport $J_c$ at 10, 20, 30 K and the magnetic fields higher than 0.1 T. For instance, Fig. 9 shows $J_c(H)$ data taken at 10 K for 50 and 100 nm films. Shown in Fig. 10 are the $J_{c,L}/J_{c,T}$ ratios extracted from $J_c(H)$ curves at 10, 20 and 30 K, like those in Fig. 9. The value $J_{c,L}/J_{c,T} \approx 1.6$ for both films is in fair agreement with MO data, but the anisotropy ratio $J_{c,L}/J_{c,T}$ at high fields only exhibits very weak field and temperature dependencies within our experimental uncertainty. However at low field (<0.12T) the anisotropy parameter $J_{c,L}/J_{c,T}$ obtained from MO (Fig. 6c) increases with increasing temperature in the 50 nm film, although the temperature dependence of $J_{c,L}/J_{c,T}$ becomes weaker for the 100 nm film and nearly flat for the 200 nm film.

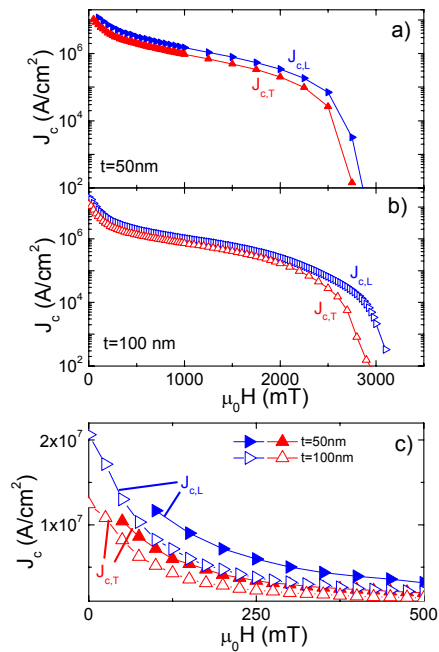

**Figure 9:** Field dependence of transport $J_c$ at 10 K in the 50 nm (a) and 100 nm (b) c-axis tilted films. In the bottom panel c), the low field regime for both films is zoomed and plotted in linear scale. Notice that $J_{c,L}$ and $J_{c,T}$ values for the 50 nm film (filled symbols) are larger than for the 100 nm film (open symbols).

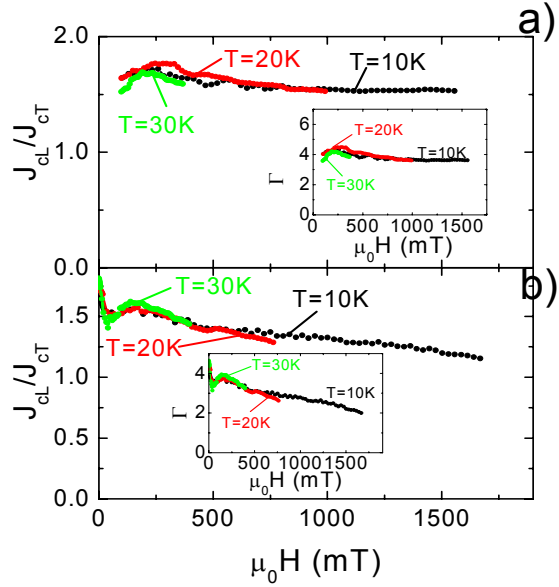

**Figure 10:** Field dependencies of the ratio of transport $J_{c,L}/J_{c,T}$ for the 50nm (a) and 100 nm (b) c-axis tilted films. Inset shows the mass anisotropy parameter $\Gamma$ extracted from Eq. (5) for both films.

Before presenting a model of pinning which accounts for the behavior of the $J_c$ anisotropy in c-axis tilted films, we summarize the main experimental results presented so far.

- Surfaces of c-axis tilted films reveal characteristic terraced structures with rugged edges, which provide strong surface pins, irrespective of the direction of the in-plane current.
- From resistivity curves measured in flat films ($\rho$) and in c-axis tilted films in the directions parallel ($\rho_L$) and perpendicular ($\rho_T$) to the terrace steps, it comes out that $\rho$ and $\rho_L$ are nearly identical, while $\rho_T$ is larger, which indicates that current transport is determined by intrinsic band parameters and that the presence of the steps allows us to separate in-plane from combined in-plane and out-of-plane contributions.
- $J_c$ from MO in both directions parallel ($J_{c,L}$) and perpendicular ($J_{c,T}$) to the terrace steps shows that the critical current decreases with increasing film thickness and $J_{c,L}$ of the 50 nm film is extremely high, reaching almost $9 \cdot 10^7$ A/cm$^2$ at low temperature.
- In the 50nm and 100 nm films $J_{c,L}$ is higher than $J_{c,T}$, while for the 200 nm film the currents are virtually equal in both directions. The anisotropy $J_{c,L}/J_{c,T}$ increases with increasing temperature and field for the 50 nm film, but these dependences become weaker with increasing thickness.
- Transport $J_c$ values measured in a perpendicular field of 0.1 T are extremely high, exceeding $10^7$ A/cm$^2$ and compatible with the MO data measured at few tens of mT fields. Also the ratio $J_{c,L}/J_{c,T} \approx 1.6$ measured by transport in the 50nm and 100 nm films is in fair agreement with MO data.

## 4. Model

The behavior of $J_c$ observed on c-axis tilted $MgB_2$ films indicates very strong pinning of vortices. Since the measured $J_c$ is of the order of the depairing current density $J_d \approx 10^8$ A/cm$^2$ [3] a significant part of the vortex core length is pinned by defects, causing the energy gain of the order of the superconducting condensation energy $B_c^2/2\mu_0$ in the correlated volume $\sim \xi^3$ [31,32], where $B_c$ is the thermodynamic critical field. Such strong pinning is usually associated with non-superconducting nanoprecipitates with dimensions close to the vortex core size. The vortex core in $MgB_2$ has a composite structure with different coherence lengths $\xi_\sigma$ and $\xi_\pi$ over which the superfluid densities in $\sigma$ and $\pi$ bands are suppressed [38]. Because the thickness of our films t = 50 - 200 nm is comparable to $\xi_\pi \approx 50$ nm at T << $T_c$ [12,13], vortices perpendicular to the film surface can be effectively pinned by only a few defects, as depicted in Fig. 11. Characteristic pinning mechanisms labeled A, B and C can be identified. Case A corresponds to pinning by a few defects in the bulk. Intermediate case B is a combination of pinning by a single defect in the film and pinning of vortex ends at the surface. The latter can be due to the nano-terraces characteristic of our films (see Fig. 1). The lateral sizes and heights of slanted terrace steps are of the order of $\xi$, which can make them strong pinning centers. In case C, $J_c$ is mostly dominated by surface pinning. Pinning mechanisms A, B and C result in different dependencies of $J_c$ on the film thickness t.

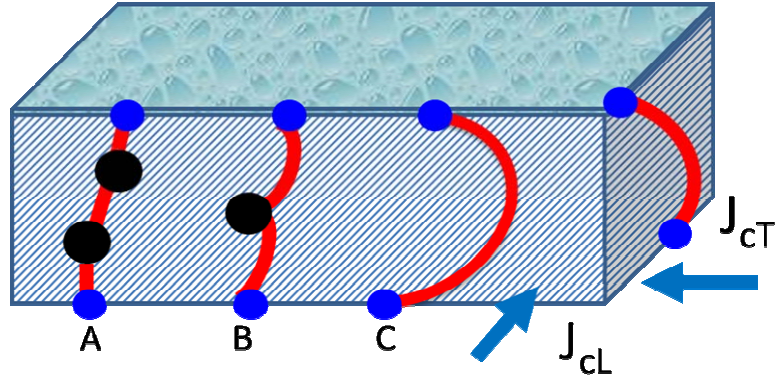

**Figure 11:** Three different types of pinning in a thin c-axis tilted film, as explained in the text. Here red lines show pinned vortex segments, black circles show pinning centers in the bulk (nanoprecipitates) and blue circles show pinning centers on the surface (related to surface nanoterraces).

Case A can be evaluated using the standard arguments of the collective pinning theory for a single vortex at low fields [32]. Depinning occurs as the net Lorentz force $F_L = t\phi_0 J$ exceeds the net pinning force $F_p$. The Lorentz force causes the vortex to bow out as shown in Fig. 11, but in a thin film, the length of vortex segments between the pins may be $\sim \xi$ so bending distortions are weak and the vortex remains nearly straight. In this case $F_p$ is proportional to the mean-square fluctuation $N^{1/2}$ of the total number of pins $N=t/\ell$ along the rigid vortex, $F_p \sim f_p r_p N^{1/2}/\ell$, where $f_p$ is the elementary pinning force, $r_p$ is the pin interaction radius, and $\ell$ is the mean pin spacing. Hence,

$$J_c \sim \frac{U_p}{\phi_0}\left(\frac{n_p}{t}\right)^{1/2} \sim J_d \left(\frac{\xi}{t}\right)^{1/2} \qquad (1)$$

where $U_p=f_p r_p$ is the pinning energy and $n_p=\ell^{-3}$ is the volume pin density and $J_d=\phi_0/3^{3/2}\pi\mu_0\lambda^2\xi$. Here we took $\ell \sim r_p \sim \xi$ and $U_p \sim \xi^3 B_c^2/\mu_0$ as was discussed above.

The inverse square root thickness dependence of $J_c(t)$ in Eq. (1) implies N>>1, which may not occur if t ~ $\ell$. The latter corresponds to the case B in Fig. 11, where $J_c$ is limited by one pin in the bulk and by surface pins, unless there are many weaker pinning defects smaller than $\xi$. For a single strong pin in the bulk, a force balance, $t\phi_0 J=f_p$ yields the inverse thickness dependence $J_c=f_p/t\phi \sim J_d \xi/t$. However, in the case of B and C, bending distortions of vortices may be essential because of significant mass anisotropy in the $\sigma$ band and the c-axis tilt in our $MgB_2$ films. Depinning for the case B and C occurs due to the pin breaking mechanism [33,34,35], in which the vortex ends are fixed at the surface by very strong pins. As J increases, the vortex bows out until the critical value $J_c$ is reached at which the vortex end segments become nearly parallel to the surface. At $J>J_c$ vortex gets depinned and starts sliding along the film as the ends of vortex segments become parallel to the surface and reconnect with antivortex images. The same mechanism works for a single pin in the middle of the film (case B): the vortex gets depinned as two vortex segments above and below the pin reconnect. Using the method of vortex-antivortex images, the case B with the pin in the middle of the film and without surface pinning reduces to the case C, resulting in the same $J_c$ for a given film thickness.

The critical current density is calculated using the anisotropic elastic theory for a single vortex [34] in which the pinned vortex segment under uniform Lorentz force has an elliptical shape stretched along the ab planes. Taking the x-axis parallel to the film surface and assuming that the c-axis is tilted by the angle $\theta$ with respect to the y-axis normal to the film, the elliptical vortex loop in which the surface tension is balanced by the Lorentz force is described by [32]:

$$(x\sin\theta + y\cos\theta)^2 \Gamma^2 + (x\cos\theta + y\sin\theta)^2 = \left(\frac{\varepsilon_0}{\phi_0 J}\right)^2 \qquad (2)$$

where $\Gamma=\lambda_c/\lambda$ is the ratio of the London penetration depths for the field along the ab plane ($\lambda_c$) and the field along the c-axis ($\lambda$), and $\varepsilon_0=\phi_0^2\ln(\Gamma t/2\xi)/4\pi\mu_0\lambda^2$ is the line energy of the vortex parallel to the c-axis. Here $J_c$ is determined by the condition that the elliptical semi-loop fits in the film, touching its upper and lower surfaces, (y = ± t/2 and $\partial y/\partial x=0$). Hence

$$J_{c,L} = \frac{\phi_0 \ln(t/2\xi_c)}{2\pi\mu_0\lambda^2\Gamma t}\left(\Gamma^2\sin^2\theta + \cos^2\theta\right)^{1/2} \qquad (3)$$

where $\xi_c=\xi/\Gamma$ is the coherence length along the c-axis. For $\theta=0$, Eq. (3) reduces to $J_c$ obtained by Brandt [34]. Notice that Eq. (3) derived from the single-band London theory can be used only for qualitative estimates, given the existence of two coherence lengths ($\xi_\sigma$, $\xi_\pi$), and the fact that our thinnest films have t~$\xi_\pi$. Yet because the dependence of $J_c$ on $\xi_c$ is weak, Eq. (3) may capture the physics of pinning in our c-axis tilted $MgB_2$ films.

Equation (3) shows that the c-axis tilt increases $J_{c,L}$ if current flows along the ab planes as the Lorentz force causes bending deformation of vortices with both in-plane and out-of-plane displacements. Because the mass anisotropy decreases the vortex line tension for soft in-plane distortions while increasing the line tension for hard out-of-plane distortions [32], the c-axis tilt stiffens vortices against bending deformations caused by in-plane currents. As a result, the longitudinal $J_{c,L}$ increases as the tilt angle $\theta$ increases. By contrast, currents flowing across tilted ab planes cause only soft in-plane vortex deformations (see Fig. 11). Thus, the transverse $J_{c,T}$ due to depinning of vortex semi-loops remains independent of θ, which can also be obtained from the anisotropic scaling theory [35]. The critical current density $J_{c,T}$ is then given by Eq. (3) with θ=0:

$$J_{c,T} = \frac{\phi_0 \ln(t/2\xi_c)}{2\pi\mu_0 \lambda^2 \Gamma t} \tag{4}$$

Equations (3) and (4) can also be used to evaluate $J_{c,L}$ and $J_{c,T}$ for a single strong pin in the middle of the film. As the film thickness increases, collective pinning effects become more important, resulting in a crossover from $J_c \propto t^{-1}$ to $J_c \propto t^{-1/2}$. In this case, $J_c$ can be roughly evaluated by replacing t with the mean length $\ell$ of pinned vortex segments and dividing $J_c$ by the fluctuation factor $N^{1/2}=(t/\ell)^{1/2}$. As a result, we arrive to Eqs. (3) and (4) in which t is replaced with the geometric mean $(\ell t)^{1/2}$. The ratio $J_{c,L}/J_{c,T}$ takes a simple form,

$$J_{c,L}/J_{c,T} = \left(\Gamma^2 \sin^2\theta + \cos^2\theta\right)^{1/2} \tag{5}$$

In this model the ratio $J_{c,L}/J_{c,T}$ depends only on the tilt angle and the electron anisotropy, and thus is temperature independent at low fields for which the London model is qualitatively applicable. For t=50-100 nm, Eqs. (3) and (4) result in $J_{c,L} \approx 8 \cdot 10^7$ A/cm$^2$ which is of the order of the nominal GL depairing current density $J_d = \phi_0/3^{3/2}\pi\mu_0\lambda^2\xi \approx 1.4 \cdot 10^8$ A/cm$^2$ if we take λ=100 nm and $\xi=\xi_\sigma$=13 nm at T<<$T_c$ [3]. At such high $J_c$, pairbreaking effects in the weak π band become essential, suppressing the superfluid density $n_\pi(J)$ and decoupling π and σ bands at $J \sim J_d^{(\pi)} = \phi_0/3^{3/2}\pi\mu_0\lambda^2\xi_\pi$ [36]. For the typical ratios $\xi_\pi/\xi_\sigma$ = 3-4, we obtain $J_d^{(\pi)} \sim 3 \cdot 10^7$ A/cm$^2$, not much higher than the measured $J_{c,L}$ shown in Fig. 9. Therefore, in our thinnest 50 nm and 100 nm films current transport may be limited by pairbreaking effects in the weak π band. This scenario is consistent with the observed temperature dependence of $J_{c,L}/J_{c,T}$, as will be discussed below.

Theory of vortex pinning taking into account pairbreaking effects in multiband MgB$_2$ at low temperatures requires numerical simulations of the two-band Eilenberger or Eliashberg equations [37,38]. Instead, we discuss qualitative manifestations of pairbreaking effects in a simple model in which the penetration depths for two weakly couple bands depend on their respective superfluid densities $n_{\sigma,\pi}$ and effective band masses $m_{\sigma,\pi}$:

$$\lambda^{-2} = \frac{e^2}{\mu_0}\left(\frac{n_\pi}{m_\pi} + \frac{n_\sigma}{m_\sigma}\right), \quad \lambda_c^{-2} = \frac{e^2}{\mu_0}\left(\frac{n_\pi}{m_\pi^{(c)}} + \frac{n_\sigma}{m_\sigma^{(c)}}\right) \tag{6}$$

where the index (c) corresponds to the effective masses along the c-axis and e is the electron charge. In the clean limit the values of $n_\pi$ and $n_\sigma$ approach the partial carrier densities in π and σ bands, respectively at T<<$T_c$. In the dirty limit the ratio $n_i/m_i$ is replaced with $N_i D_i \Delta_i \tanh(\Delta_i/2T)$ where i = (σ,π), $N_i$ is the partial density of states and $D_i$ is the diffusion coefficient [38,39,40,41,42]. The anisotropy parameter is then:

$$\Gamma^2(J) = \frac{\dfrac{n_\pi}{m_\pi} + \dfrac{n_\sigma}{m_\sigma}}{\dfrac{n_\pi}{m_\pi^{(c)}} + \dfrac{n_\sigma}{m_\sigma^{(c)}}} \cong \frac{m_\pi^{(c)}}{m_\pi} + \frac{n_\sigma m_\pi^{(c)}}{n_\pi(J)m_\sigma} \tag{7}$$

Notice that the key contribution of a nearly isotropic π band makes Γ not very large even in the limit of a 2D σ band ($m_\sigma^{(c)} >> m_\sigma$). Here current pairbreaking mostly affects $n_\pi(J)$ if J is not too close to $J_d$, where the dependence of $n_\pi(J)$ on J increases can be calculated using the two-band GL theory [43]. Reduction of $n_\pi(J)$ as J increases manifests itself in the increase of the effective $\lambda(J)$ and $\Gamma(J)$ in Eqs. (6) and (7) as compared to the London model. In turn, these effects reduce $J_{c,L}$ and $J_{c,T}$ in Eqs. (3) and (4) making them closer to the experimentally observed values. This model suggests that the anisotropy parameter $J_{c,L}/J_{c,T}$ should increase in thinner films because they have higher $J_c$ and thus larger $\Gamma(J)$ due to stronger suppression of $n_\pi$ by current pairbreaking effects.

## 5. Discussion

The thickness dependence of $J_c(t)$ indicates whether bulk ($J_c \sim 1/\sqrt{t}$) or surface ($J_c \sim 1/t$) pinning dominates. As follows from the analysis presented above, $J_c(t)$ can be written as a sum of surface and bulk pinning contributions

$$J_c(t) = \frac{A}{t} + \frac{C}{\sqrt{t}} \qquad (8)$$

where A and C quantify the strengths of surface and bulk pinning, respectively. To separate these two contributions, we re-plot $J_c(t)$ extracted from the MO experimental data, in the form of $J_c \cdot t$ as a function of $t^{1/2}$. According to Eq. (8), such graph should give a straight line with the slope C and the intersect A with the vertical axis. The results shown in Fig. 12 indicate that Eq. (8) qualitatively describes the experimental $J_c(t)$ data, which clearly have both surface and bulk pinning contributions represented by the first and the second terms in Eq. (8). The upshift of the straight lines for the tilted films (larger parameter A) as compared to the flat films in Fig. 12 suggests that the surface pinning is stronger in tilted films, consistent with extra pinning by the surface steps discussed above. It is also evident that the upshift for $J_{c,T}(t)$ is smaller than for $J_{c,L}(t)$, in agreement with Eq. (5) which describes the anisotropy of $J_c$ caused by tilted ab planes. These results show that in thinner c-axis tilted films pinning the effect of surface terrace steps becomes more pronounced. However, there is not much room for many strong pins per vortex given that the size of the vortex core diameters $2\xi_\sigma \approx 26$ nm and $2\xi_\pi \approx 100$ nm [12,13,44] are of the order of t in the 50 nm films. In these films there may be only one or two pins at the surface and the substrate interface, plus one or no strong bulk pins per vortex line, although there may be many smaller randomly distributed bulk pinning defects on scales of few nm.

Given the complicated microstructure masked by strains shown in Fig. 2, we cannot unambiguously identify the main defects responsible for bulk pinning. Moreover, the use of Eq. (8) obtained under the assumption of $t \gg \xi_\pi$ for the films with $t \cong \xi_\pi$ may require a phenomenological replacement of t with $t + t_0$ in the denominators to cut off the divergence at t = 0 so that $J_c$ does not exceed $J_d$ in thin films if $t_0 \cong \xi_\pi \cong 50$ nm. We can also expect reduction of A due to suppression of superconductivity at the film surface due to lattice mismatch between the film and the substrate and oxides at the surface [45]. Moreover, the change of the nanostructure of surface steps for different film thicknesses shown in Fig. 1 brings about additional mechanisms of the thickness dependence of $J_c(t)$. Indeed, for the thinnest 50 nm film, the terrace size is close to the vortex core diameter, which provides the most effective surface pinning. As the thickness increases, the surface steps coalesce and become smaller, while the terrace sizes L increase, making the surface pinning less effective and increasing the dominant role of bulk pinning at self field. However, at higher fields at which the intervortex spacing $a = (\phi_0/B)^{1/2}$ is of the order of the terrace size L, surface pinning may become more efficient. For L = 100 nm, the matching field is $B_m = \phi_0/L^2 = 0.2$ Tesla. As a result, the thickness dependence of $J_c(t)$ can vary for different magnetic fields. Temperature can also affect the parameter A in Eq. (8) given that $\xi_\pi(T)$ and $\xi_\sigma(T)$ increase as T increases, so the vortex core size matching characteristic sizes of terrace cells at low temperatures may become larger at higher T, resulting in the decrease of surface pinning of vortices as temperature increases. Yet despite these materials complexities of real $MgB_2$ films, our simple pinning model appears consistent with our experimental data and describes, at least qualitatively, the thickness dependence of $J_c$ both parallel and perpendicular to the surface steps in tilted films.

Another test of the model comes from the analysis of $J_c$ anisotropy in tilted films. Insets in Fig. 10 show the behavior of the anisotropy parameter $\Gamma$ obtained from in-field transport $J_c$ data using Eq. (5). The $\Gamma$ values extracted from the high-field data shown in Fig. 10 are around 4 and are nearly independent of temperature, as expected from our model. In Fig. 13, we summarize the field, temperature and thickness dependence of $\Gamma$. Here Fig. 13a shows $\Gamma(H)$ extracted from MO and transport data of the tilted 50 nm film at low fields, while in Fig. 13b shows the temperature dependencies $\Gamma(T)$ extracted from MO measurements for all three films. The anisotropy parameter $\Gamma$ exhibits several features indicating a significant role of current pairbreaking in the $\pi$ band. Indeed, the value of $\Gamma = 4$ for the 50 nm film is about 2-3 times higher than the low-field ratio $\Gamma = \lambda_c/\lambda_{ab} = 1.3-2$ extracted from the measurements of $H_{c1}$ on $MgB_2$ single crystals [46,47]. This increase of $\Gamma(J)$ in Eqs.(7)-(9) is a natural consequence of very high $J_c$ resulting in stronger suppression of superconductivity in the nearly isotropic $\pi$ band and thus amplification of the contribution of the highly anisotropic $\sigma$ band. The same effects also result in the increase of $\lambda^2(J)$ in Eqs. (3), (4), and (6). As it was shown earlier, Eq. (4) yields $J_{c,T} = 8 \cdot 10^7$ A/cm$^2$ for the low-field values $\Gamma = 2$ and $\lambda = 100$ nm. The current-induced depletion of superconducting condensate in the $\pi$ band increases $\Gamma$ by a factor of 2, so according to Eqs. (6) and (7), $\lambda$ should also increase by a factor $\approx 2$. These two effects reduce the London estimate of $J_{c,T}$ in Eq. (4) by a factor of 8, bringing $J_{c,T}$ close to the observed value of $10^7$ A/cm$^2$ shown in Fig. 9.

The above scenario has several consequences, which can be checked experimentally. First, the pairbreaking effects in the $\pi$ band should diminish in thicker films as $J_c(t)$ in Eq. (8) decreases. This trend is indeed observed in our films, as shown in Fig. 13b. Clearly, $\Gamma$ decreases from 4 for t = 50 nm to 3 for t = 100 nm to its conventional low-field value $\approx 1.2$ [46,47] at t = 200 nm. The temperature dependence of $\Gamma$ shown in Fig. 13b is also consistent with different temperature dependences of $n_\sigma(T)$ and $n_\pi(T)$ in Eq. (7) where $n_\pi(T)$ is reduced by strong current pairbreaking in the thinnest 50 nm film above 20 K. Another manifestation of the pairbreaking effects in the $\pi$ band is the significant increase of $\Gamma(H)$ and $\lambda(H)$ as H increases and approaches the "virtual upper critical field" $H_\pi = \phi_0/2\pi\xi_\pi^2 = 0.1-0.3$ T [12,13] above which the normal vortex cores in the $\pi$ band overlap. A clear indication of this effect, which follows from Eqs. (6) and (7) is visible in Fig. 7c and 13a. This is also consistent with is a significant (by a factor ~10) decrease of $J_c(H)$ below 0.4 T clearly seen in Fig. 9. Yet the overlap of the $\pi$ vortex cores at $H > H_\pi$ does not completely suppress the superfluid density in the $\pi$ band which remains finite due to the intrinsic proximity effect caused by interband interaction with the main $\sigma$ band. Because the isotropic contribution of $\pi$ band persists even at $H > H_\pi$, the anisotropy parameter $\Gamma$ defined by Eq. (7) increases, but not to the level defined by the mass anisotropy of the $\sigma$ band.

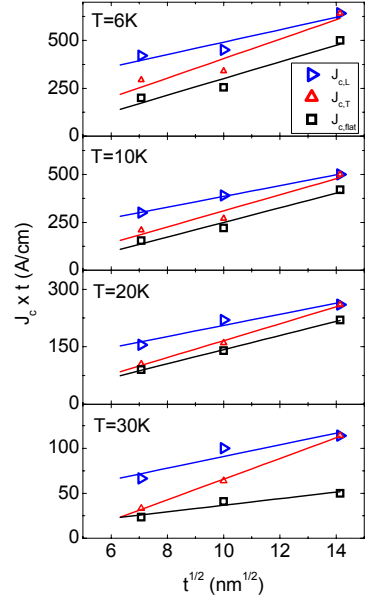

**Figure 12:** Thickness dependencies of $J_{c,L}$, $J_{c,T}$ and $J_{c,flat}$ extracted from MO data at different temperatures. Data are plotted in the form of $J_c t = A + C\sqrt{t}$ in order to separate bulk and surface pinning contributions, as discussed in the text.

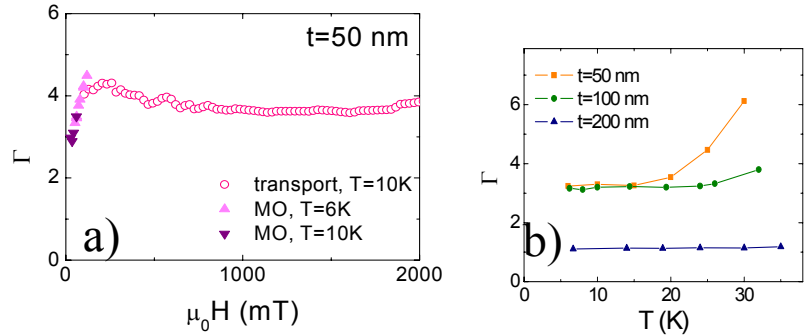

**Figure 13: a)** Field dependence of the anisotropy parameter $\Gamma$ extracted from MO and transport measurements on the 50 nm c-axis tilted film at low temperature, using Eq. (5). **b)** Temperature dependencies of the anisotropy parameter $\Gamma$ for extracted from MO measurements on films of different thickness.

In addition to the above intrinsic mechanisms, the change in the surface step nanostructure shown in Fig. 1 may also contribute to the anisotropy of $J_c$. Indeed, for the 50 nm film, the lateral distribution of surface steps is nearly isotropic, while thicker films have surface terrace structures with visible preferential orientation which can affect the ratio of $J_{c,L}/J_{c,T}$. Notice that the behavior of the anisotropy ratio $J_{c,L}/J_{c,T}$ observed on our c-axis tilted $MgB_2$ films is rather different from in c-axis tilted $YBa_2Cu_3O_{7-\delta}$ films in which $J_{c,L}/J_{c,T}$, decreases with temperature [18], indicating different mechanisms of $J_c$ anisotropy in these two cases. In $YBa_2Cu_3O_{7-\delta}$ films, the $J_c$ anisotropy has been attributed to anti-phase boundaries at the edges of the miscut steps that are effective pinning centers for $J_{c,L}$ and not for $J_{c,T}$ [22]. However, the HR-TEM of our c-axis tilted $MgB_2$ films show the existence of subgrain boundaries. On the other hand grain boundaries in $MgB_2$ are nearly transparent to current so they may not be as strong pinning centers and current-limiting defects as they are $YBa_2Cu_3O_{7-\delta}$, [48]. Hence the relationship between microstructure and pinning behind the behavior of c-axis tilted $YBa_2Cu_3O_{7-\delta}$ films is different in c-axis tilted $MgB_2$ films in which $J_{c,L}/J_{c,T}$ is determined by intrinsic band contributions.

## 6. Conclusion

In this work we performed MO and transport measurements of critical current density $J_c(T,H)$ in c-axis oriented and c-axis tilted epitaxial $MgB_2$ films to separate intrinsic contributions of π and σ bands. We observed a clear correlation of the in-plane anisotropy of $J_c(T,H)$ with micro and nanostructures with for 50 and 100 nm c-axis tilted films imaged by high resolution transmission and scanning electron microscopy. The HR-TEM and SEM investigations of the c-axis tilted films revealed terrace step structures on the surface with lateral sizes of the order of the vortex core diameter, which suggest that such steps can effectively pin vortices. We observed very high $J_c$ of the order of the depairing current density, and the anisotropy of $J_c$ in the c-axis tilted films in which $J_{c,L}$ flowing parallel to the ab planes is larger than $J_{c,T}$ crossing the tilted ab planes. The measured field and temperature dependencies of $J_c$ along the directions either parallel or perpendicular to the surface steps can be attributed to anisotropic effective band masses, and significant suppression of superfluid density in the π band due to current pairbreaking in 50 and 100 nm films. The analysis of the thickness dependence of $J_{c,L}$ and $J_{c,T}$ shows that the enhancement of $J_c$ observed on tilted c-axis films mostly results from extra pinning of the surface terrace steps. Our results suggest that the out-of-plane critical current controlled by π band could limit $J_c(T,H)$ in randomly oriented polycrystals.

**Acknowledgements**

The work at the National High Magnetic Field Laboratory (USA) was supported by the National Science Foundation under NSF/DMR1157490 and by the State of Florida. The work at Old Dominion University was supported by the USA Department of Energy under Grant No. DE-SC0010081. The work at University of Tokyo was partially supported by the Japan Science and Technology Agency, PRESTO. The work at Temple University was supported by the USA Department of Energy under Grant No. DE-SC0004410. The authors are grateful to David C. Larbalestier for useful discussions and suggestions.